\documentclass{easychair}

\usepackage{doc}
\usepackage{booktabs}
\usepackage{cite}
\usepackage{hyperref}

\title{Proof Recommendation System for the HOL4 Theorem Prover}

\author{
Nour Dekhil
\and
Adnan Rashid
\and
Sofiène Tahar
}

\institute{Department of Electrical and Computer Engineering\\
Concordia University, Montreal, QC, Canada \\
\email{\{n\_dekhil,rashid,tahar\}@encs.concordia.ca}}

\begin{document}

\maketitle

We introduce a proof recommender system for the HOL4 theorem prover~\cite{HOL4}. Our tool is built upon a transformer-based model~\cite{transf} designed specifically to provide proof assistance in HOL4. The model is trained to discern theorem proving patterns from extensive libraries of HOL4 containing proofs of theorems. Consequently, it can accurately predict the next tactic(s) (proof step(s)) based on the history of previously employed tactics. The tool operates by reading a given sequence of tactics already used in a proof process (in our case, it contains at least three tactics), referred to as the current proof state, and provides recommendations for the next optimal proof step(s).


Figure 1 depicts the major steps taken to develop the proof recommendation tool. The initial block (highlighted in blue) refers to the construction of a HOL4 proofs dataset. In the dataset construction phase, we are abstracting the proof scripts to only include the tactics used to prove a theorem or a lemma. This process involves initially identifying the theorems or lemmas within each \texttt{sml} file, followed by recording the tactics used to prove each one of them.

\vspace*{-3mm}

\begin{figure}[http]
\centering
  \includegraphics[width=0.85\textwidth]{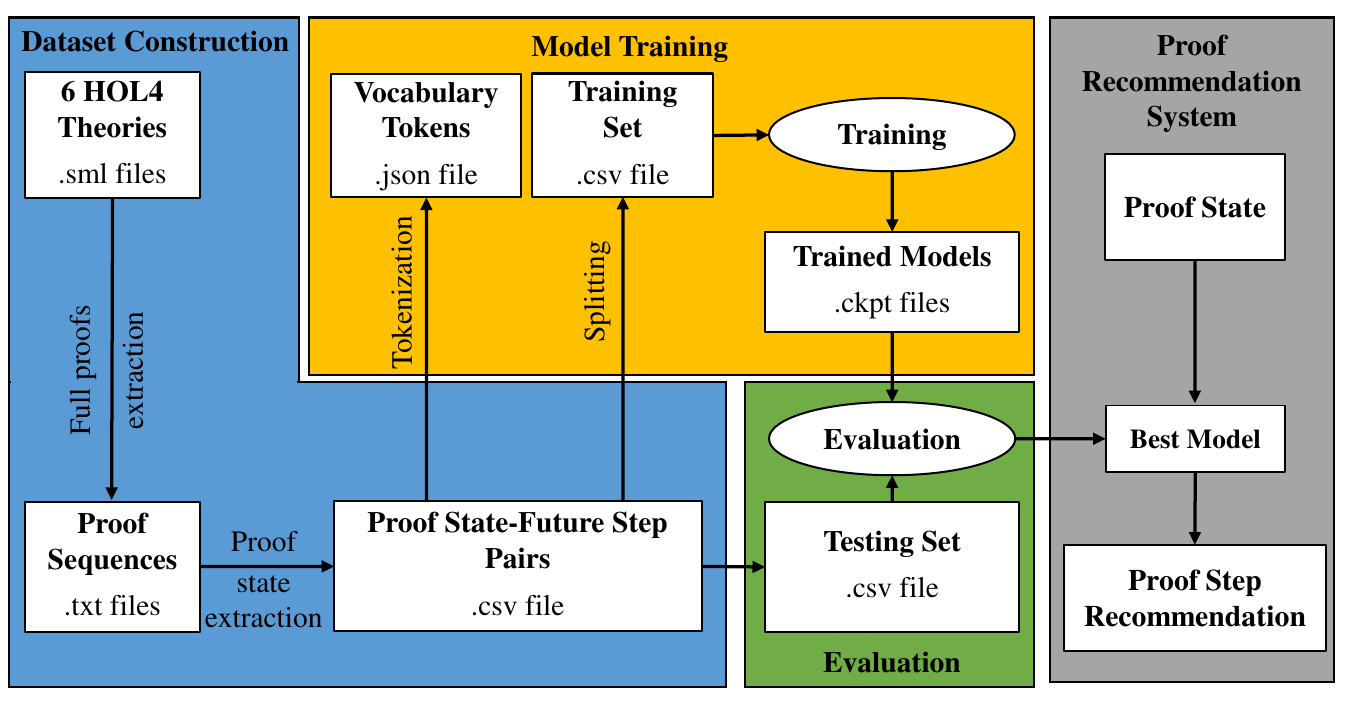}
  \label{fig:example12}
  \vspace{-4mm}
\caption{Proof Recommendation System.}
\vspace{-6mm}
\end{figure}

\vspace{4mm}

We created large proof sequences datasets (Datasets $1$-$5$) from five HOL4 theories \cite{dataset1,dataset2,dataset3,dataset4,dataset5} developed by the Hardware Verification Group (HVG) of Concordia University alongside an already available dataset created using the real arithmetic theory of HOL4 (Dataset $6$)~\cite{dataset6}. For experimental purposes, we combined all datasets into Dataset $7$. Our objective is to predict the subsequent tactic from a sequence of previously employed tactics. To accomplish this, we approach this challenge as a multi-label classification task using language models. To facilitate this, we restructure the dataset into pairs of current proof states and possible future tactics. More details on the datasets used for classification are given in Table~\ref{tab:mytable}. \\

\begin{table}[http]
\vspace*{-8mm}
\caption{Summary of the used Datasets}
\vspace*{2mm}
    \centering
    \resizebox{\linewidth}{!}{%
    \begin{tabular}{lccccccc}
        \toprule
        & Dataset 1 & Dataset 2 & Dataset 3 & Dataset 4 & Dataset 5 & Dataset 6 & Dataset 7 \\
        \midrule
        \textbf{Distinct Tactics} & 115 & 132 & 26 & 44 & 32 & 89& 162\\
        \textbf{Proofs} & 1,873 & 2,475 & 153 & 295 & 61 & 279 & 5,136\\
        \textbf{Proof States} & 43,167 & 57,602 & 2,973 & 7,371 & 1,784 & 3,259& 116,156\\
        \bottomrule
    \end{tabular}%
    }
    \label{tab:mytable}
    \vspace{-8mm}
\end{table}
\vspace{5mm}
We experimented with various transformer-based language models, such as BERT~\cite{bert}, RoBERTa~\cite{roberta}, and T5~\cite{t5} for these datasets to identify the most effective model based on our evaluation. After splitting the restructured datasets into a $90$-$10$ ratio for training and testing, we proceeded to train the selected models (block highlighted in yellow) using a grid search of hyperparameters optimization. Given the multitude of possible tactics available at each proof state, we chose to provide multiple recommendations for the next proof step. To assess the accuracy of these recommendations (block highlighted in green), we use the \textit{n}-correctness rate, which measures the likelihood that a correct tactic from the testing dataset is among the top-\textit{n} recommended tactics, where \textit{n} signifies the number of recommended tactics evaluated against the correct tactic. We found out that RoBERTa demonstrated superior performance across most cases for $n = 7$. As a result, we deploy it into our proof recommendation tool (block highlighted in grey).

With the aim of efficiently predicting the next tactic ($k = 1$, where $k$ represents the number of future tactics to predict) for the majority of theory datasets, we also challenged our tool by attempting to predict two future tactics. Table~\ref{tab:mytable22} provides further details of the experimental results for RoBERTa in predicting one future tactic ($k = 1$) and two future tactics ($k = 2$). After examining the performance results across different datasets, it seems that the variations arise from the diversity and patterns unique to each dataset, as well as the range of tactics employed. Specifically, Datasets $1$-$5$ exhibit a uniformity in their proof structures, originating from one application project written by a single person, thus making the proofs more homogeneous and consistent in style. However, Dataset $6$, came from HOL4 libraries containing a diverse range of theorems regarding different mathematical concepts, presents proofs with heterogeneous patterns, making them challenging to predict. Additionally, we observed a decrease in performance when attempting to predict two future tactics, which may be attributed to the expansive space of possibilities and resulting in increased uncertainty.

\begin{table}[http]
\vspace*{-8mm}
\caption{Correctness Rates of RoBERTa Considering Top-$7$ Recommendations}
\vspace*{2mm}
    \centering
    \resizebox{\linewidth}{!}{%
    \begin{tabular}{lccccccc}
        \toprule
        & Dataset $1$ & Dataset $2$ & Dataset $3$ & Dataset $4$ & Dataset $5$ & Dataset $6$ & Dataset $7$ \\
        \midrule
        \textbf{k = 1} & 73.6\% & 79.5\% & 94.4\% & \textbf{97.8\%} & 97.6\% & 64.3\% & 89.8\%\\
        \textbf{k = 2} & 54.3\% & 58.6\% & 88.1\% & \textbf{96.8\%} & 92.2\% & 29.4\% & 80.3\% \\
        \bottomrule
    \end{tabular}%
    }
    \label{tab:mytable22}
    \vspace{-8mm}
\end{table}
\vspace{5mm}
In the recent past, several studies have integrated artificial intelligence into theorem prover tools (e.g., PVS and Coq), particularly for predicting future-proof steps. For instance, in the study reported in~\cite{yeh2023coprover}, accuracies ranging from $50\%$ to $70\%$ were achieved for the top $3$-$5$ recommendations, while the work in~\cite{blaauwbroek2020tactic} achieved $87\%$ accuracy for the top $3$, and the one in~\cite{luan2021using} reported $54.3\%$ accuracy for the top $10$. In comparison, our tool surpasses results reported in these studies, achieving accuracies of $77.3\%$, $89.88\%$, and $93.7\%$ for the top $3$, $7$, and $10$ next tactic recommendations, respectively, measured on the combined Dataset 7. The current tool version is available to try online~\cite{hol4prs}. In the future, we plan to expand it to include more HOL4 theories and enhance its interfacing with HOL4. In addition, we are investigating its potential to automatically generate complete proofs, considering the need for optimization given the exponential growth in combination possibilities with the proof sequence length. To address this, we plan to use some advanced tree search algorithms.

\bibliographystyle{unsrt}
\bibliography{easychair.bib}
\end{document}